\documentclass[showpacs,floatfix]{revtex4}
\usepackage{graphicx,psfrag,amsmath,amssymb,amsfonts,latexsym,color,epsf,dcolumn,graphpap}

\definecolor{red}{rgb}{1,0,0}
\definecolor{blue}{rgb}{0,0,1}
\definecolor{skyblue}{rgb}{0,0,.5}
\definecolor{green}{rgb}{0,1,0}
\definecolor{orange}{cmyk}{0,.4,1,0}

\begin{document}
\title{Numerical approach to simulating interference phenomena in a two-oscillating mirrors cavity }
\author{Paula I. Villar$^{a}$, Alejandro Soba $^b$, Fernando C.~Lombardo$^a$}
\affiliation{
$^a$Departamento de F\'\i sica {\it Juan Jos\'e
Giambiagi}, FCEyN UBA and IFIBA CONICET-UBA, Facultad de Ciencias Exactas y Naturales,
Ciudad Universitaria, Pabell\' on I, 1428 Buenos Aires, Argentina.\\
$^b$CNEA - CONICET 
Centro At\' omico Constituyentes; Av. Gral. Paz 1499, San Mart\' in, Argentina}
\date{today}

\begin{abstract}
\noindent 
We study photon creation in a cavity with two perfectly conducting moving mirrors.
We derive the dynamic equations of the modes and study different situations concerning various movements of the
walls, such as translational or breathing modes.
We can even apply our approach to one or three dimensional cavities and reobtain well known results of cavities with one moving mirror. We compare the numerical results with analytical predictions and discuss the effects of the intermode coupling in detail as well as the 
non perturbative regime. We also study the time evolution of the energy density as well and provide analytic justifications for the different 
results found numerically. 
\end{abstract}
\maketitle
\section{Introduction}\label{sec:intro}

The mechanical interaction between a moving mirror and the radiation field has always been an interesting issue of study. This is not only due to its practical purposes but also to its representation of a fundamental system in quantum optics. Besides the change of the zero point energy of the quantum vacuum provoked by static boundary conditions a second, there is a  yet even more fascinating feature of the quantum vacuum arising  when considering dynamical boundaries conditions. The presence of moving boundaries leads to a non stable vacuum electromagnetic state, resulting in the generation of real photons, which is an amazing demonstration of the existence of quantum vacuum fluctuations of quantum electrodynamics (QED), referred to in the literature as the dynamical Casimir effect (DCE) \cite{reviews} or motion-induced radiation. DCE is a common name ascribed to the processes in which photons are generated from vacuum due to the external time variation of boundary conditions for some field \cite{12,13,15}. For the usual electromagnetic case this corresponds to the fast motion of a mirror or modulation of the dielectric properties of the mirror/ intra-cavity medium \cite{16}. 
 
Research in the field has mainly concentrated on one-dimensional models and a few works with more realistic three dimensional models \cite{fulling,dodonov1,mundarian,dodonov2,Ruser}. Since the amount of radiation generated is very small, much attention has been paid to the study of one-dimensional model for which the effect is enhanced, as for example through the parametric resonance condition. The main difference between one and three-dimensional cavities is that, while in one dimension the cavity's frequency spectrum is equidistant and leads to strong intermode interactions, in three dimensions the spectrum is in general non equidistant, and only a few modes may be coupled \cite{crocce,dodonov3}.
A cavity made of two perfectly parallel reflecting mirrors, one of which oscillates with a mechanical frequency equal to a multiple of the fundamental of the static cavity (while the other one is at rest) is a typical case where the mentioned enhancement takes place. In most works this problem has been analytically studied through a perturbative expansion of the equations of motion of the field in terms of the small oscillation amplitude to find an approximative solution at short times. Although the direct measurement of radiation generated by moving mirrors is an important experimental challenge, it was asserted \cite{Wilson} that photon creation induced by time-dependent boundary conditions has been observed experimentally in superconducting circuits. This experiment consists of a coplanar waveguide terminated by a superconducting quantum interference device (SQUID), upon which a time-dependent magnetic flux is applied. A related experiment involving a Josephson metamaterial embedded in a microwave cavity has been described in Ref. \cite{Paraoanu}. These experiments stimulated new theoretical research on role of dynamical Casimir physics in quantum  information processing, quantum simulations and engineering of nonclassical states of light and matter \cite{19,20,21,22,23}. There are also ongoing experiments aimed at measuring the photon creation induced by the time-dependent conductivity of a semiconductor slab enclosed by an electromagnetic cavity \cite{Bragio}, as well as proposals based on the use of high frequency resonators to produce the photons, and ultracold atoms to detect the created photons via superradiance \cite{Kim}. 

Recent studies have indicated that DCE could be implemented even using a single two-level atom (qubit) with time-dependent parameters, such as the transition frequency or the atom-field coupling strength \cite{5, 25,26,27,28}. Generation of excitation from vacuum occurs due to the counter-rotating terms in the Rabi Hamiltonian, which for many years had been neglected under the Rotating Wave Approximation. On the other side, 
it has been proved that an ensemble of two-level atoms collectively coupled to the electromagnetic field of a cavity, driven at low frequencies and close to a quantum phase transition, stimulates the production of photons from the vacuum. This paves the way to an effective simulation of the DCE \cite{vedral}. 

The case of cavities with two moving mirrors has also been considered due to a perturbative treatment \cite{chinos} and also using a technique inspired in the renormalisation group method, where the solution to the set of generalised Moores' equations is valid both for short and long times, improving perturbative approaches \cite{DalvityDiego}. Authors considered two mirrors oscillating resonantly at the same frequency, allowing for different amplitudes and dephasing between mirrors in one dimensional cavities.
As we shall see, the radiation induced strongly depends on the relation among the amplitudes, the frequency and the different phase in the walls oscillations. As predicted in \cite{DalvityDiego}, we show that for some relations among the variables, there is constructive interference which leads to an exponential growth of particles inside the cavity. For some other relations, there is destructive interference and hence no vacuum radiation. We also show that our solution accounts for other physical solutions (non perturbative regime for example) as for one oscillating wall.

In this paper we will present a detailed numerical analysis of the particle creation rate, along with analytical considerations on 
the cases mentioned above.
The paper is organised as follows. In Sec. \ref{bc} we shall present the equation of motion for the field modes of the electromagnetic field. In Sec. \ref{onewall}, we begin by analysing the photon creation in the case where one wall of the cavity is at rest and the other one oscillates with a multiple frequency of  the fundamental of the static cavity. Therein, we shall consider a three dimension as well as one dimensional cavities.  In Sec. \ref{2wall}, we concentrate on the case of two oscillating mirrors, one at each end of the cavity. We focus on the study of different cases of dephasing between the oscillating walls, i.e. zero and $\pi$-dephasing movements. Finally, in Sec.\ref{conclusiones} we make our conclusions.

\section{Boundary Conditions}
\label{bc}

We consider a rectangular cavity formed by perfectly conducting walls with dimensions $L_x$ ($L_y$ and $L_z$ as well if we consider a three dimensional cavity).  The mirrors placed at $x=L(t)$
and $x=R(t)$ are at rest for $t<0$ and begin to move at $t=0$ following a given trajectory $L(t)$ and $R(t)$ respectively. 
We assume these trajectories as prescribed for the problem and that they work as a time-dependent boundary condition for the field. 

We start with the field operator $A(x,t)$ for the vector potential which satisfies the wave equation $\Box \vec{A}=0$. In terms of the creation 
$a_k^{\dagger}$ and annihilation $a_k$
operators, the field operator can be expressed as:
\begin{equation}
A(x,t)= \sum_k^{\infty} \bigg[ \hat{a}_k \psi_k (x,t) + \hat{a}_k^{\dagger} \psi_k^*(x,t) \bigg].  
\end{equation}

In the previous equation, $\psi_k(x,t)$ are the mode functions of the field and are chosen so as to satisfy the boundary conditions,  i.e. $\psi_k(L(t),t)=0$ and $\psi_k(R(t),t)=0$.
We shall firstly consider the instantaneous mode basis, for the $1+1$ field \cite{comentario}:
\begin{equation}
\phi_k(x,t)=\sqrt{\frac{2}{R-L}} \sin \left[\frac{k \pi (x-L)}{R-L} \right],
\end{equation}
and write each mode as:
\begin{equation}
\psi_k(x,t)=\sum_{m=1}^\infty Q_{k m}(t) \phi_m (x,t),
\end{equation}
where $m$ is a positive integer. 

By considering small amplitude motions of the walls, we can write their trajectories as given by:
\begin{equation}
L(t) = A_L \epsilon_L \sin(\Omega_L t + \phi_L)
\end{equation}

and
\begin{equation}
R(t)=L_0 - A_R \epsilon_R \sin(\phi_R) + A_R \epsilon_R \sin (\Omega_R t + \phi_R),
\end{equation}
where $L_0$ is the cavity length in the static situation, and $\epsilon_R$, $\epsilon_L$ 
are small (dimensionless) parameters which characterise the small deviations of the walls from the initial static positions, and 
$A_L$ and $A_R$ are amplitudes. The mirrors can oscillate in phase or not, depending on the values of $\phi_L$ and $\phi_R$.
By inserting  the expansion of field modes into the wave equation and integrating over spatial dimensions leads to the equation of motion for the canonical variables  expressed as:

\begin{widetext}
\begin{eqnarray}
\ddot {Q}_m ^{(n)} + \omega^2_m(t)  Q_m^{(n)} &=& \frac{1}{L_0}\sum_s b_{ms} \dot{Q}_s^{(n)} +\frac{1}{4 L_0^2} \sum_s \bigg
\{ g_{ms}  + (3 (\dot R - \dot L)^2 + 2 L_0 (\ddot R- \ddot L) )  a_{ms} +
 (r_{ms}^1+ r_{ms}^2) \bigg\}  Q_s^{(n)}, 
\label{ecmov}
\end{eqnarray}
\end{widetext}
where, in the general 3+1 dimensions case, 
$\omega_m(t)= \sqrt{\frac{(m \pi)^2}{(R(t)-L(t))^2}+  k_{\parallel}^2}$ is the mode frequency. The coefficients in the equation 
are defined as follows: 

\begin{widetext}

\begin{equation*}
 a_{ms}=  \left\lbrace
  \begin{array}{l}
     1 \text{ if } m = s \\
     0 \text{ otherwise} \\
  \end{array}
  \right.
\end{equation*}

\begin{equation*}
 b_{ms}=  \left\lbrace
  \begin{array}{l}
     \frac{4 m s }{(m^2-s^2)} (\dot R (-1)^{m+s}-\dot L) \text{ if } m \neq s \\
     0 \text{ otherwise} \\
  \end{array}
  \right.
\end{equation*}

\begin{equation*}
 g_{ms}=  \left\lbrace
  \begin{array}{l}
     \frac{m s }{(m^2-s^2)} \left[24 \dot L (\dot R- \dot L)-24 \dot R (\dot R-\dot L) (-1)^{m+s}+8 L_0 ( \ddot R(-1)^{m+s}- \ddot L\right]
 \text{ if } m \neq s \\
     3 (\dot L - \dot R)^2)-2 (L_0 (\ddot R - \ddot L)  \text{ if } m=s \\
  \end{array}
  \right.
\end{equation*}

\begin{equation*}
 r_{ms}^1=  \left\lbrace
  \begin{array}{l}
     \frac{16 m^3 s }{(m^2-s^2)^2} \dot L (\dot R- \dot L)(-1+(-1)^{m+s}) \text{ if } m \neq s \\
     2 m^2 \pi^2 \dot L (\ddot R + \ddot L) \text{ if } m=s \\
  \end{array}
  \right.
\end{equation*}

\begin{equation*}
 r_{ms}^2=  \left\lbrace
  \begin{array}{l}
     \frac{-16 m^3 s }{(m^2-s^2)^2} ( \dot R-\dot L) [\dot L+(\dot L-2 \dot R)(-1)^{m+s} ]\text{ if } m \neq s \\
     \frac{2}{3} (\dot R -\dot L) (3 (\dot L-\dot R)+ m^2 \pi^2 (\dot R + \dot L)) \text{ if } m=s \\
  \end{array}
  \right.
\end{equation*}
\end{widetext}

\section{Numerical Method}

In this Section we describe the numerical method used for solving the equation of motion of the field modes determined by  Eq.(\ref{ecmov}). 
In order to solve the equation of motion of the $n$ modes, we perform a change of variables  in order to obtain a new system of equations:
\begin{eqnarray}
\dot{Q}_m &=& U_m, \nonumber \\
\dot{U}_m &=& -\omega_m^2(t) Q_m + \sum_s S_{m s} (t) Q_{s}, 
\label{modelo}
\end{eqnarray}
where $S_{ms}(t)$ is a bracket proportional to $Q_s$  in Eq.(\ref{ecmov}). 
We have dropped the supra-index for simplicity. 
The initial conditions, specified for each field mode in all cases are:
\begin{equation}
Q_k^{(n)}(0)=\frac{1}{\sqrt{2 \omega_n} }\delta_{k,n}; ~~~~~~\dot{Q}_k^{(n)}(0)= -i \sqrt{\frac{\omega_n}{2} }\delta_{k,n};
\end{equation}
which indicate that the field modes and their derivatives are continuous at $t=0$, as long as $R(t)$, $L(t)$, $\dot{R}(t)$ and $\dot{L}(t)$ are smooth functions.
For a time dependence of the boundary, either $R(t)$ or $L(t)$, which is not sufficiently smooth, (for example discontinuities in its time-derivative),  one may expect spurious particle creation.

We have used an integration scheme based on a fourth order Runge-Kutta-Merson numerical method between $t=0$ and a maximum time $t_{\rm max}>0$.
In all cases, the moving walls are at rest at $t=0$ and then, the perturbation is turned on for times between $0<t<t_F$, with $t_F<t_{\rm max}$, where the walls remains static again (this can be applied to either one or two moving walls).  For times $t<0$ and $t> t_F$, the cavity is a static one and we know the set of orthonormal functions. The quantisation of the system is straightforward through creation and annihilation operators:
\begin{eqnarray}
Q_n (t<0) &=&  \frac{1}{\sqrt{2 \omega_n}} ( \hat{a}_n e^{-i \omega_n t} + \hat{a}_n^{\dagger} e^{i \omega_n t} ), \label{freemode}
\end{eqnarray}
with frequency  $\omega_n(t)=1/L_0\sqrt{(\pi n)^2 + k_{\parallel}^2}$, where $L_0$ is the initial length of the cavity and $k_{\parallel}$ is associated to the non-dynamical dimensions of de cavity ($L_y$ and $L_z$). 
The time-independent annihilation and creation operators $\hat{a}_n$, $\hat{a}^{\dagger}_ n$ associated with the particle notion for $t \leq 0$ are subject to the commutation relations $[\hat {a}_n, \hat {a}_m]=[\hat {a}_n^{\dagger}, \hat {a}_m^{\dagger}]=0$ and  $[\hat {a}_n, \hat {a}_m^{\dagger}]=\delta_{nm}$. The initial vacuum state $|0,t \leq 0 \rangle$ is defined by:
\begin{equation}
a^{\dagger}_{n} |0,t \leq 0 \rangle =0 ~~~\forall ~~n.
\end{equation}

When the cavity dynamics is switched on at $t=0$ and the walls follow the prescribed trajectory $L(t)$ and $R(t)$, the field modes are coupled. Then, the $Q_n$ can be written as:
\begin{equation}
Q_n(t \geq 0) = \sum_m \frac{1}{\sqrt{2\omega_m}} ( \hat{a}_m \epsilon_n^m(t) + \hat{a}_m^{\dagger} \epsilon_n^*(t)), 
\end{equation}
with complex functions $\epsilon_n^m(t)$ that satisfy the equation of modes. When the motion ceases and the walls are at rest again for $t>t_F$, $Q_n(t)$ can be expressed again as:
\begin{eqnarray}
Q_n (t \geq t_F) &=&  \frac{1}{\sqrt{2 \omega_n^{1}}} ( \hat{A}_n e^{-i \omega_n^1(t-t_F)} + \hat{A}_n^{\dagger} e^{i \omega_n^1 (t-t_F)} ),
\end{eqnarray}
with $ \omega_n^{1} = \omega_n^{1}(t \geq t_F)$ and the annihilation and creation operators $\hat{A}_n$ and $\hat{A}_m^{\dagger}$  corresponding to the particle notion for
$t \geq  t_F$. The final vacuum state $|0, t \geq t_F \rangle$ is defined by:
\begin{equation}
A^{\dagger}_{n} |0,t \leq t_F \rangle =0 ~~~\forall ~~n.
\end{equation}

As expected, the initial state particle operators $\hat{a}_n$ and $\hat{a}_n^{\dagger}$ are linked to the final state operators $\hat{A}_n$ and $\hat{A}_n^{\dagger}$ by a Bogolubov transformation $\hat{A}_n= \sum_m (A_{mn}(t_F) \hat{a}_m + B_{mn}^*(t_F) \hat{a}_m^{\dagger} )$.
 The total number of particles created  in a mode $n$ during the motion of the wall is given by the expectation value of the particle number operator $\hat{A}_n^{\dagger} \hat{A}_n$ associated with the particle notion for $t \geq t_F$ with respect to the initial vacuum state: 
\begin{equation}
N_n(t_F)=\langle 0, t\leq |A^{\dagger}_n A^{\dagger}_n |0, t\leq 0 \rangle = \sum_m |B_{m n }(t_F)|^2.
\label{Nparticulas}
\end{equation}

In order to obtain the numerical results presented in the following Sections we proceed in the following way. Two cut-off parameters $\Lambda$ (for the field modes considered) and  $\Lambda_m$ (for the number of canonical variables considered) are introduced to make the system of differential equations finite and suitable for a numerical treatment. The system of $n \times m$ coupled differential equations is then evolved numerically from $t = 0$ up to a final time $t_F$ and the expectation value of Eq.(\ref{Nparticulas}) is calculated for several times in between. By doing so we interpret $t_F$ as a continuous variable such that Eq.(\ref{Nparticulas}) becomes a continuous function of time. Consequently, the stability of the numerical solutions with respect to the cut-offs has to be ensured. In particular $\Lambda$ will be chosen such that the numerical results for the number of particles created in single modes are stable. In most cases, it is enough to choose $\Lambda_m=\Lambda$.
In our units, the spectral modes $k_n=\Omega_n$ are given in units of $1/L_0$ ($k_n L_0$ is dimensionless) and consequently time is measured in units of $L_0$. 

%

\section {One moving mirror}
\label{onewall}

We can start by studying the photon creation when only one wall is moving following $R(t)$ and the other one is at rest in $x=0$ (for example, by setting  $\epsilon_L=0$ and $\epsilon_R \neq 0$). In this case, we can take any value of $\phi_R$, say $\phi_R=0$. If we excite the cavity with an external frequency $\Omega_R$ such that $\Omega_R= 2 \omega_1$, we shall produce parametric resonance induced by the moving mirror at $x=R(t)$.  We can consider either a one dimensional cavity or a three dimensional one, since this approach can be applied to either $1+1$ or $3+1$ dimensions, by taking into account that in the latter case, the ``moving walls" are in the x-direction, while the field satisfies Dirichlet boundary conditions in the other walls of the cavity 
($y$ and $z$ directions).  The important difference between one- and higher-dimensional cavities is that the frequency spectrum in only one spatial dimension is equidistant while it is in general non-equidistant for more spatial dimensions. An equidistant spectrum yields strong intermode coupling whereas in case of a non-equidistant spectrum only a few or even more modes may be coupled allowing for exponential photon creation in a resonantly vibrating three-dimensional cavity.
For both Dirichlet boundary conditions,  the eigenfrequencies inside the cavity satisfy the following condition:
 \begin{equation}
 \omega_n=\frac{1}{L_0}\sqrt{(\pi n)^2 + M^2}, 
 \label{dirichletfrequencies}
 \end{equation}
 where $n$ is natural number and we have set $M^2 \equiv {k_\parallel}^2$. If the field is massless (which corresponds to a one dimensional cavity), then the spectrum is equidistant, i.e. the difference between two consecutive eigenfrequencies is constant. Otherwise, if $M$ has arbitrary nonzero values, the spectrum is non-equidistant, corresponding to the one of a three-dimensional cavity.

\subsubsection{Three-Dimensional cavity}

As we have  stressed at the end of the previous Section, when considering a three dimensional cavity, the parallel component of the wave number $k_{\parallel}=\pi \sqrt{(n_y/L_y)^2+ (n_z/L_y)^2}$ can be associated with the non-dynamical cavity dimensions,  can be identified with the ``mass" of a massive field \cite{Ruser}. Consequently, the number of TE-mode photons created in a three-dimensional cavity equals the number of scalar particles of mass $k_{\parallel}$ created in a one-dimensional cavity. Then,  we perform the simulations by considering $\Omega=\Omega_R= 2 \omega_1$ and different values of a ``mass" $M$ for simulating the particle creation in a three-dimensonal cavity.
\begin{figure}[h!]
\begin{minipage}{8.5cm}
\begin{center}
\includegraphics[width=7.5cm]{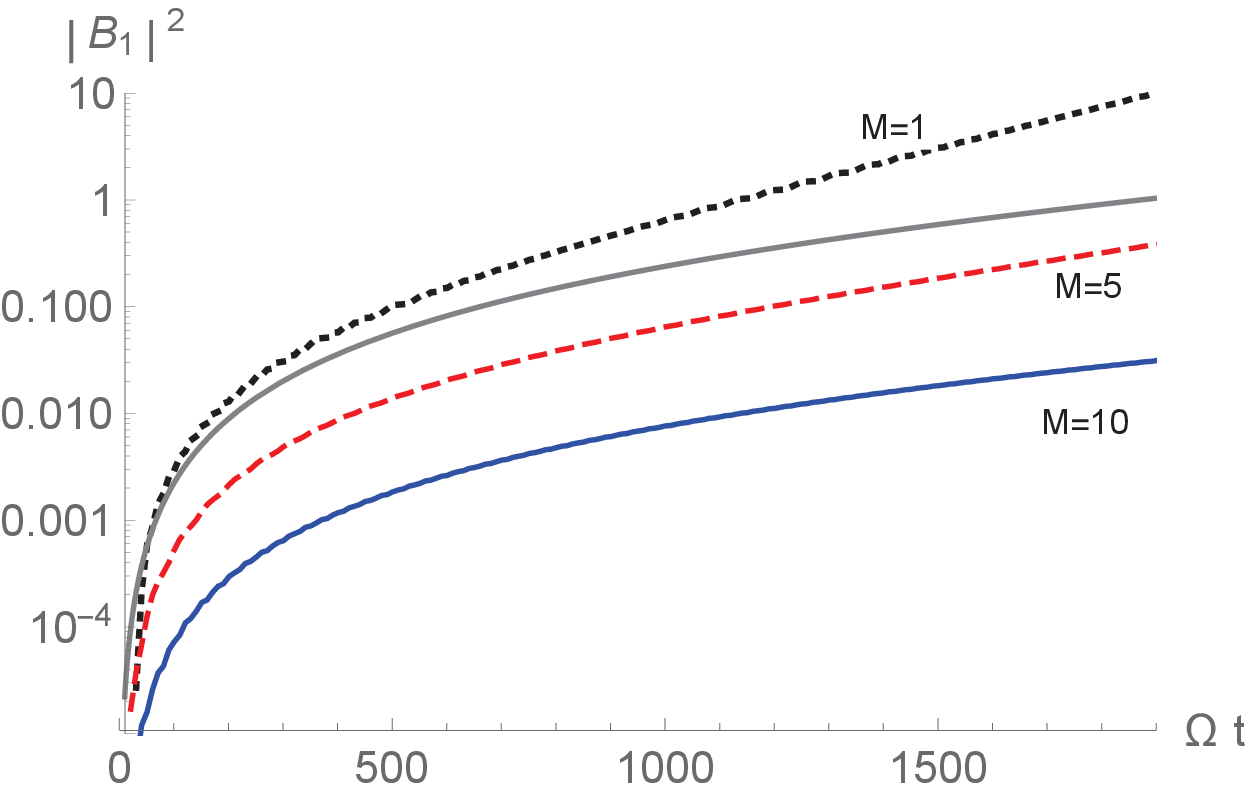}\end{center}
\end{minipage}
\begin{minipage}{8.5cm}
\begin{center}
\includegraphics[width=7.5cm]{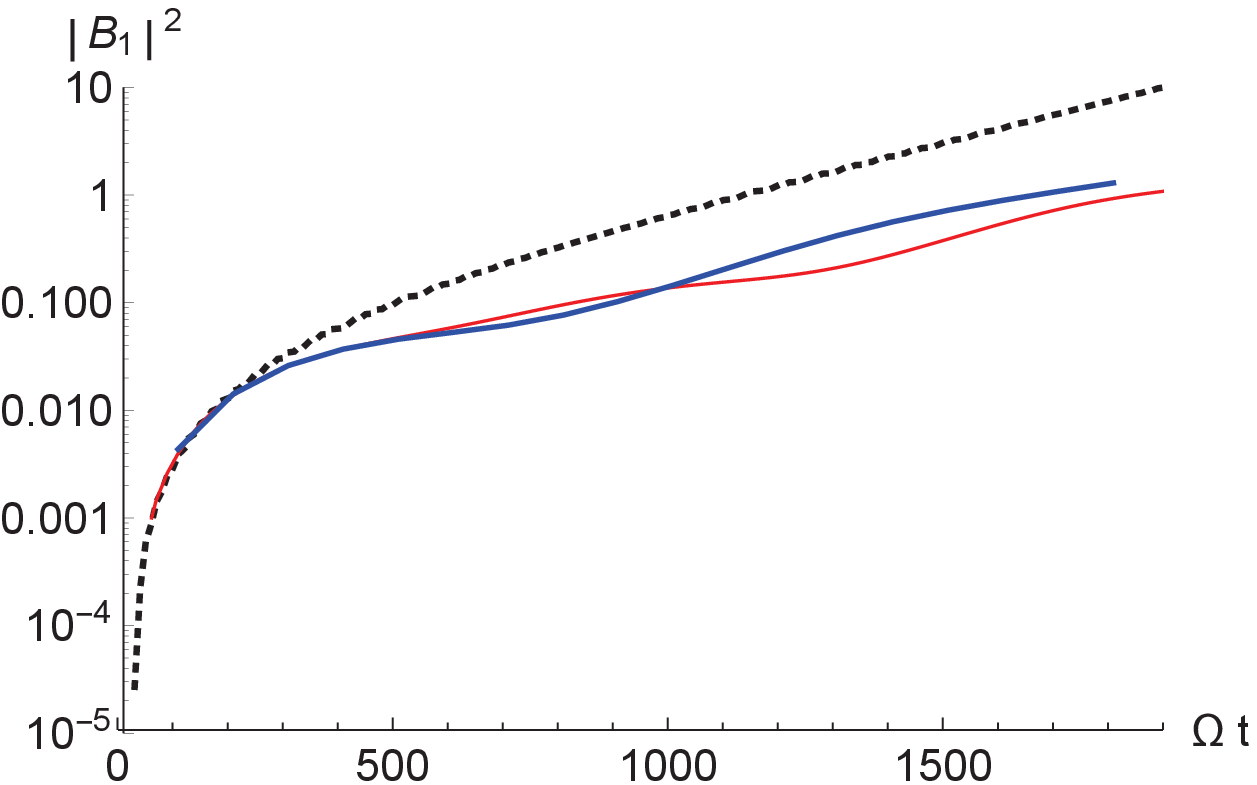}
\end{center}
\end{minipage}
\caption{(Color online) (a) Left: Behaviour of the $\vert{B}_1\vert^2$ coefficient as a function of the dimensionless time, for  different values of $M$ under the perturbation of $\Omega= 2 \omega_1$. The grey solid (upper) curve represents the  analytical solution for the number of particles created when $M=10$, while the blue solid (lower) curve is the numerical solution obtained for the right-moving mirror. (b) Right: Behaviour of the $\vert {B}_1\vert^2$ coefficient as a function of the dimensionless time, for  smaller values of $M$, i.e. $M=0.01$, $M=0.05$ and $M=1$.
Parameters used: $\epsilon=0.001$, $\Lambda=10$. }
\label{caso1pared}
\end{figure}
In Fig.\ref{caso1pared}, we can see the behaviour of the $\vert {B}\vert^2$ coefficient of the mode $n= 1$ of the field for different values of $M$, say  $M=1$, $M=5$ and $M=10$. It 
is expected that the particle creation would be exponential as the frequency spectrum becomes more non-equidistant. This is achieved, more evidently,  for bigger values of  $M \gg \pi/L_0$. 
In a resonant vibrating cavity, the number of TE-mode photons created in the resonant mode $n$ increases exponentially in time as Ref \cite{Ruser}:
\begin{equation}
N_n(t) = \sinh \bigg(\frac{n^2 \pi^2 \epsilon_R t}{2\Omega L_0^2}\bigg)^2.
\label{ruser}
\end{equation}
 We can see in Fig.\ref{caso1pared} the expected exponential behaviour as the mass increases. The grey solid curve (parallel to $M=10$) represents $N_1(t)$ as computed for $M=10$ in \cite{Ruser} using Eq.(\ref{ruser}). We can also see that the behaviour is similar though in our cavity it is attenuated by the multiplicative factor included in the relationship between the number of created particles $N_n$ and the coefficient $\vert B_n\vert^2$. As we decrease the value of $M$, we reach a region of values where we can neglect the mass term inside the frequency definition and obtain an equidistant spectrum. In Fig.\ref{caso1pared} (b) we can see that the behaviour induced by the resonant frequency $\Omega= 2 \omega_1$ when the $M \rightarrow 0$ changes considerably and can not be fitted by Eq.(\ref{ruser}) any longer.

\subsubsection{One dimensional cavity}

\begin{figure}[h!]
 \begin{minipage}{8.5cm}
\begin{center}
\includegraphics[width=7.5cm]{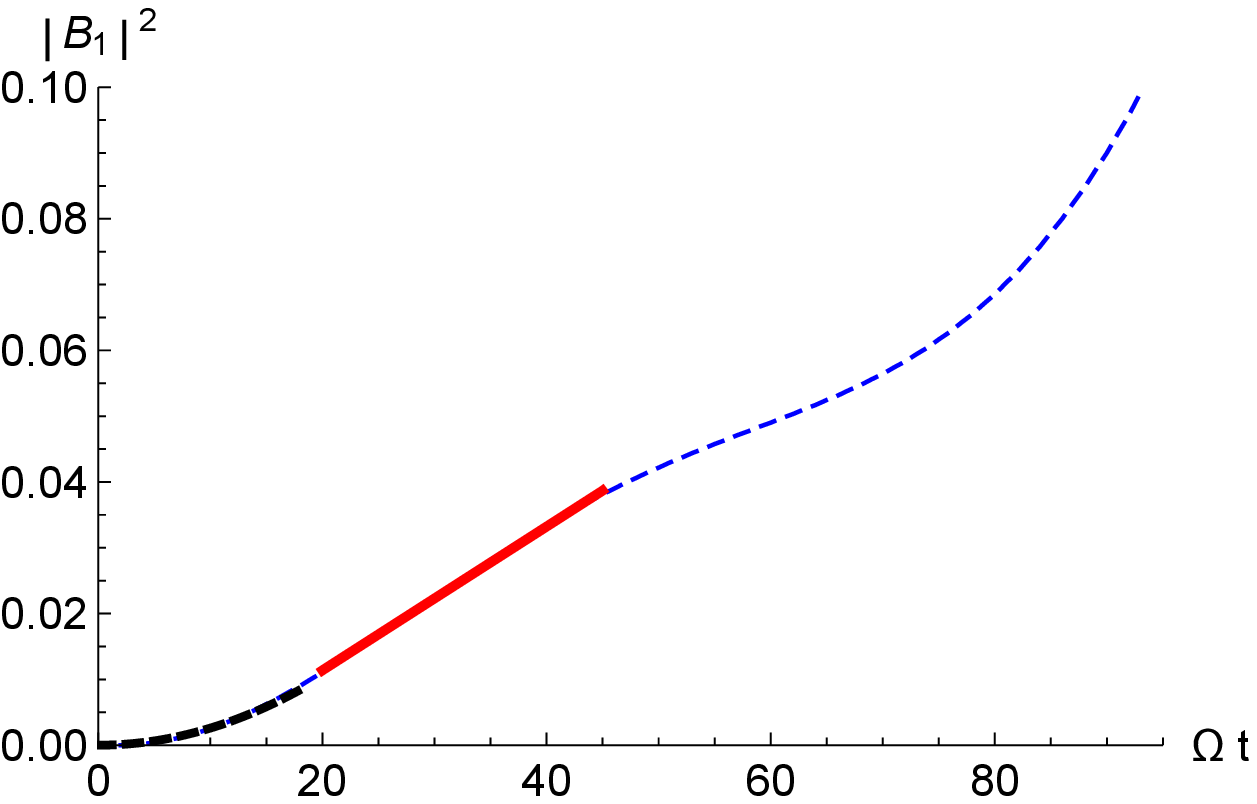}
\end{center}
\end{minipage}
\begin{minipage}{8.5cm}
\begin{center}
\includegraphics[width=7.5cm]{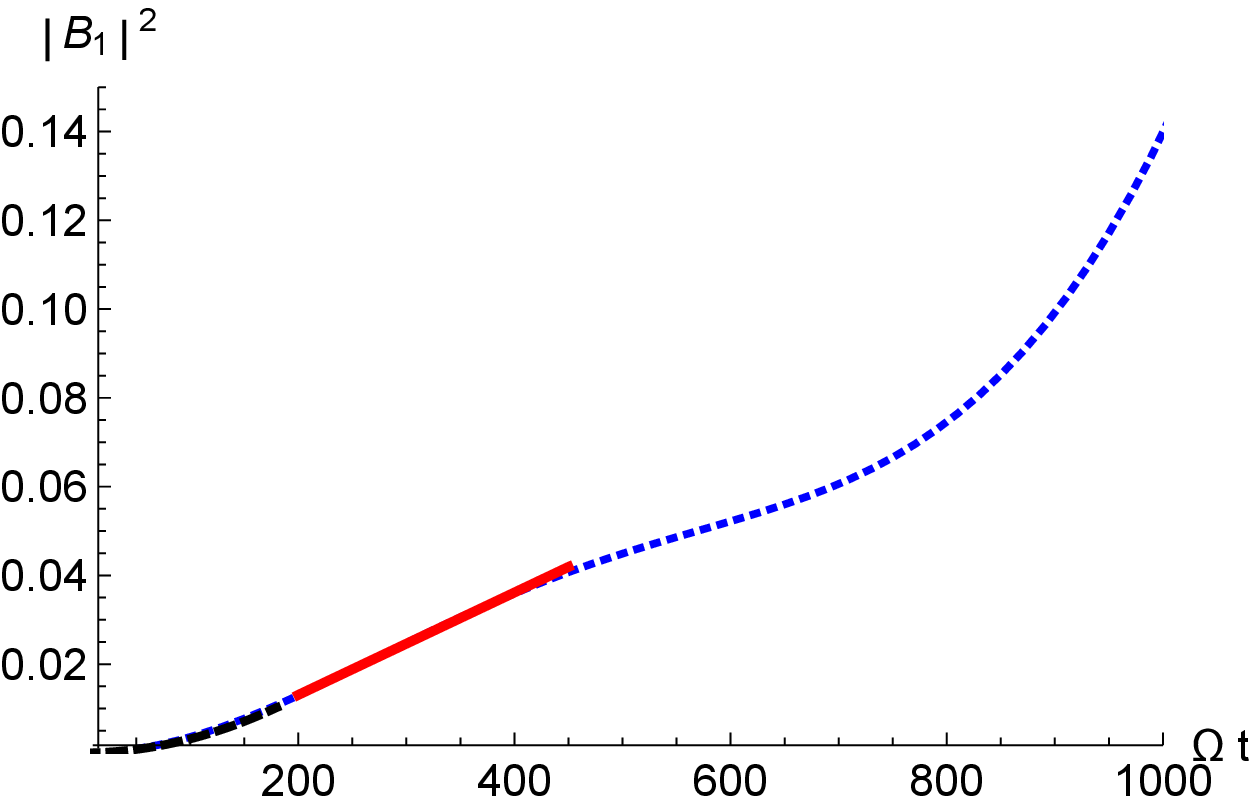}
\end{center}
\end{minipage}
\caption{(Color online) In both plots, we can see the behaviour of the $\vert {B_1}\vert^2$ coefficient of the mode $\omega_1$ of the field as a function of the dimensionless time.  For times smaller than $1/\epsilon$, the behaviour can be fitted by a quadratic curve, while for times $\Omega t \sim 1/\epsilon$ with a linear in time curve. For very long times, we find an exponential behaviour as shown in Ref. \cite{nosotros}. (a) Right: we use $\epsilon_R=0.01$ and.  (b) Left: we set  $\epsilon_R=0.001$. Parameters used: $M=0$, $\Lambda=10$. }
\label{caso1paredM001}
\end{figure}

When studying the case $M = 0$ the numerical results should converge towards the well known results for the massless  case where all modes are coupled. In this case, we know that an eigenfrequency $k$ is defined by $\omega_n = (n \pi/L_0)$. Then, if we excite the system with $\Omega= 2 \omega_1$, we should obtain a quadratic behaviour for the field mode at times $\Omega t << 1/\epsilon$. For a later time, we should expect a linear behaviour. In Fig.\ref{caso1paredM001}, we present the behaviour of $\vert {B}_1\vert^2$ for a value of $\epsilon =0.01$. For very short times, we can observe a quadratic behaviour and then a linear one. However, for values bigger than $\Omega t \gg 1/\epsilon$ we can see that the growth becomes exponential. The exponential behaviour for longer times is equivalently to what has been found in Ref. \cite{nosotros}. The coupling to a big number of field modes derives in an exponential growth at longer times, in the very non-perturbative regime.

\begin{figure}[h!]
\begin{center}
\includegraphics[width=7.5cm]{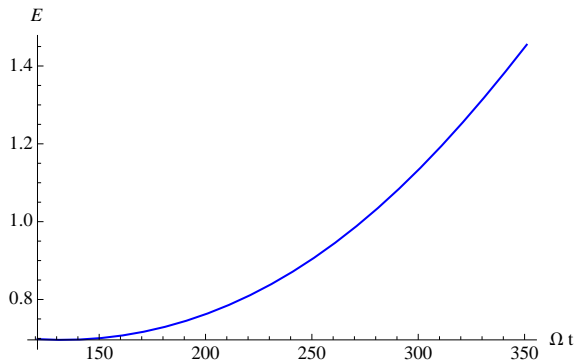}
\caption{(Color online) Energy density for a moving wall located at $x=R(t)$ as a function of the dimensionless time. The density of energy grows quadratically with time 
in the one dimensional case of a one moving mirror. Parameters: $\epsilon=0.001$, $\Lambda=10$. In our units, energy is measeured in units of $1/L_0$.}
\label{energia1pared}
\end{center}
\end{figure}

For equidistant spectrum, we can see that the coupling between an infinite number of modes leads to a quadratic growth in time of  the number of  at short timescales ($\sim 1/(\Omega\epsilon_R)$) and a linear one in the long time limit ($\sim 1/(\Omega^2\epsilon_R^2)$) as expected.
In Fig. 2(a) and 2(b) we show the dependence upon $\epsilon$ of the quadratic and linear regimes. For example $\epsilon_R=0.001$, we can observe the same behaviour but in a different temporal scale, as shown in Fig. \ref{caso1paredM001}(b). 
We can even compute the energy density for the case of a moving mirror for a one-dimensional field. In this case, the frequency spectrum is equidistant and the energy grows quadratically for short times, as in Fig.\ref{caso1paredM001}. In Fig.\ref{energia1pared}, we show the energy behaviour for an $\epsilon_R=0.001$.

\begin{figure}[h!]
\begin{center}
\includegraphics[width=8.cm]{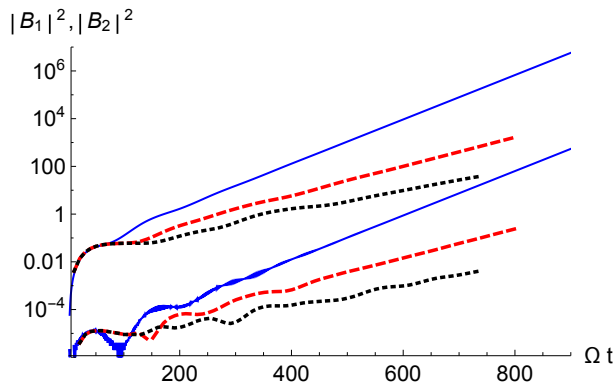}
\caption{(Color online) In this plot, we have used $\epsilon=0.01$, $\Lambda=10$ for the blue solid curves, $\Lambda=50$ for the red dashed curves and $\Lambda=100$ for the black dotted ones. In the upper values of the $\vert B\vert^2$ axis we plot the $\vert B_1\vert^2$ for the mode $n = 1$ of the field, while in the lower part of the axis is the $\vert B_2\vert^2$ of the field  as functions of the dimensionless time. We need to use a big number of modes $\Lambda$ in order to get total convergence into the final values. This plots show the dependance with $\Lambda$ and tendency to exponential growing of created particles.}
\label{comparacionmodos}
\end{center}
\end{figure}

In order to show that the behaviour at very long times is much enhanced by the finiteness of the number of modes, we show the comparison for different values of  $\Lambda$ in Fig.\ref{comparacionmodos}. Therein, we can see in a solid blue line the simulation for a cavity containing 10 field modes. The red dashed line representes a 50-modes field cavity while in the dotted black line, the cavity contains 100 modes. The upper curves represent the $\vert {B}_1\vert^2$ coefficient, while the slower ones are for $\vert{B}_2\vert^2$.  It is evident that at short times, all cavities yield the same results, but at very long times it is wiser to consider a bigger number of modes. However the qualitative behaviour remains the same, we obtain a lower exponential growth as the number of modes considered increases.
In this Section we have mainly proved numerically previous results and investigated some regions beyond the analytical studies. In the following Section we will analyse the more general case of two moving mirrors.

\section {Two moving mirrors}
\label{2wall}

In Ref.\cite{DalvityDiego} an unified analytic treatment of the dynamical Casimir effect in a one dimensional resonantly oscillating cavity for arbitrary amplitude and dephasing has been presented. Therein, it has been shown that for certain cases there is destructive interference and no radiation is generated. For others, there is constructive interference and motion-induced photons appears. When this takes 
place, the way the energy and number of created photons inside the cavity grow in time depends on the relation among  several parameters. For certain motions the growth of the energy density is exponential and for some others it is a power law. In this Section, we shall simulate the photon creation of two moving walls, for different situations. We will assume the case where $M \ll (n \pi)/(R-L)$ in order to re-obtain the results presented in Ref.\cite{DalvityDiego}, i.e. a one dimensional cavity. We will also consider the three-dimensional case by  including a mass term in the frequency of the field in a non-perturbative treatment.

\subsubsection {Non-particle induced modes}
For a particular case of equal amplitudes $\epsilon_R=\epsilon_L$ and excitation frequency $\Omega_R=\Omega_L= \omega_n$, where $n$ indicates the mode field, the energy inside the cavity oscillates  around the static Casimir value and there is no motion induced  radiation. This was first reported in \cite{chinos}, for a perturbative treatment. It can be showed that for an even value of $n$ and $\phi_R=0$ or odd value of $n$ and $\phi_R=\pi$,
\begin{figure}[h!]
\begin{minipage}{8.5cm}
\begin{center}
\includegraphics[width=8cm]{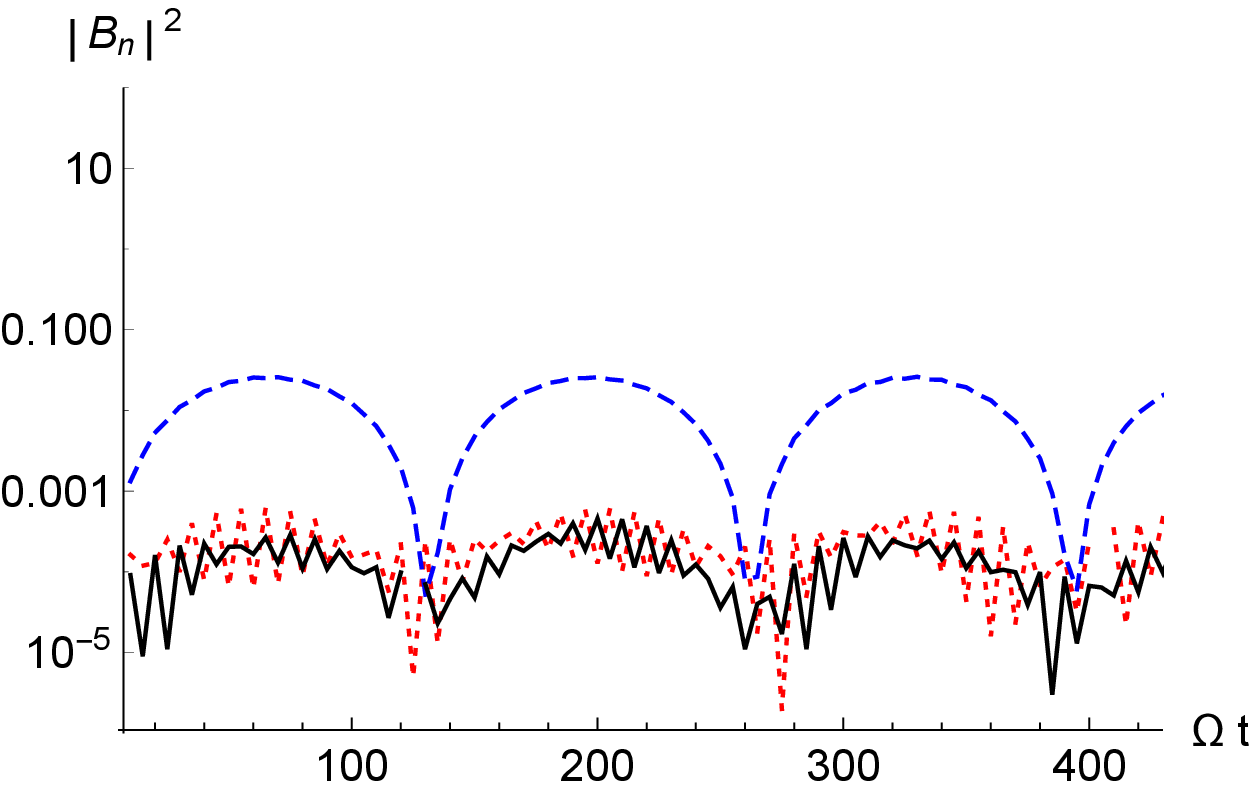}
\end{center}
\end{minipage}
\begin{minipage}{8.5cm}
\begin{center}
\includegraphics[width=8cm]{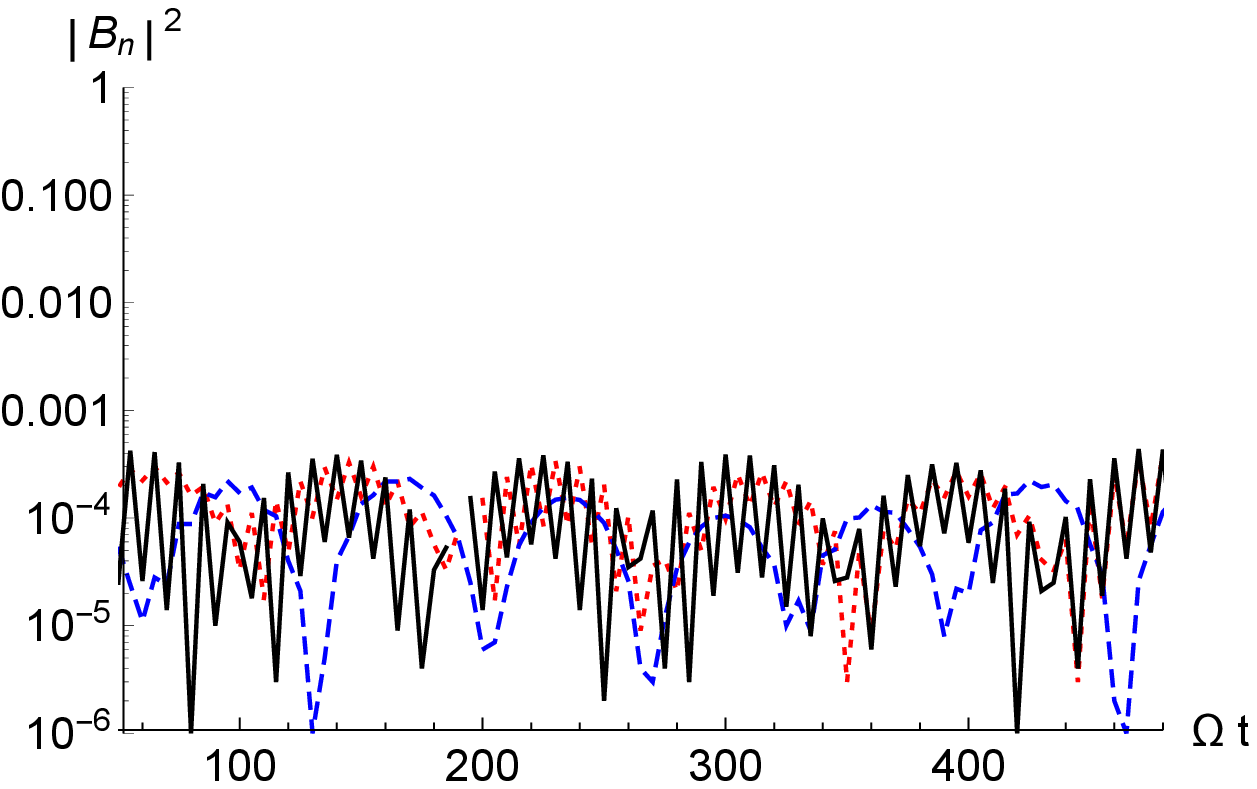}
\end{center}
\end{minipage}
\caption{(Color online) (a) Left: $\vert{B}\vert^2$ coefficient  as a function of the dimensionless time, for different field modes: $\vert{B}_3\vert^2$ (red dotted line), $\vert{B}_5\vert^2$ (blue dashed line) and $\vert{B}_7\vert^2$ (blue solid line) under $\Omega_R=\Omega_L= \omega_5$ and $\phi_R=\pi$.  
 (b) Right: $\vert {B}\vert^2$ coefficient for different field modes: $\vert{B}_3\vert^2$, $\vert{B}_4\vert^2$ and $\vert{B}_5\vert^2$ for $\Omega_R=\Omega_L= \omega_4$ and $\phi_R=0$. 
Parameters used: $\phi_L = 0$, $\epsilon=0.01$, $\Lambda=10$. }
\label{noseexcitan}
\end{figure}
 there is destructive interference among the two-moving mirrors \cite{DalvityDiego} (assuming $\phi_L=0$ in all cases). This particular behaviour is represented in 
Fig.\ref{noseexcitan} (a) and (b), where we show destructive interference and no creation of particles. 

\subsubsection {Dephased mirrors}

\begin{figure}[h!]
\begin{center}
\includegraphics[width=8.cm]{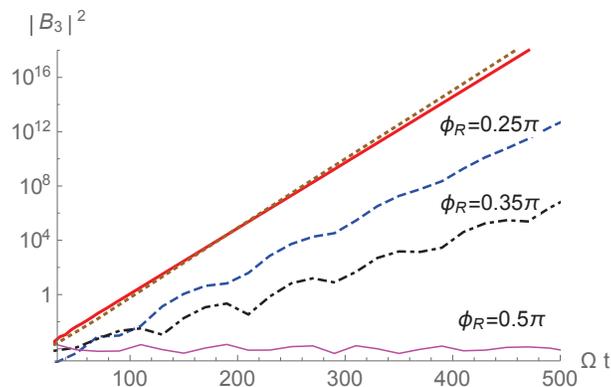}
\caption{(Color online) We present the behaviour of an odd field mode $|B_3|^2$ as a function of the dimensionless time, for different values of $\phi_R$ under a perturbation $\Omega=\omega_5$. In the case of $\phi_L= \phi_R=0$ (dotted grey line), we see there is another slope (solid red line) which corresponds to a dephasing of $\phi_L=\phi_R = \pi/4$. Both cases are examples of a translational oscillation mode,  a case which is considered separately in the following Section. In the blue dashed line we present $\phi_L = 0$ and $\phi_R = \pi/4$ and $\phi_L = 0$ and $\phi_R = 0.35 \pi$ in the dot-dashed black line. Both examples show exponential grow with time of the number of particles created with excitation $\Omega = w_5$ for mode $n= 3$. Parameters used:$\epsilon=0.01$, $\Lambda=10$.}
\label{comparacionphis}
\end{center}
\end{figure}

In this Section we follow the study of the cavity with two moving mirrors, analysing the case where the mirrors move in dephased trajectories. In particular, in Fig. \ref{comparacionphis} we show different cases in which particle creation, for odd modes, depends on the different values of the phase $\phi_R$ for a fixed value of $\phi_L = 0$. In this general case, the number of created particles grows exponentially with time for dephased motion and oscillates non-exponentially for the particular case of $\phi_L = 0$ and $\phi_R = \pi/2$ (solid pink line in Fig. \ref{comparacionphis}), when the excitation is $\Omega = \omega_5$. This case is in agreement with the result obtained in the previous Section since, it is equivalent to the total destructive interference example where there is no particle creation.  We have also included cases of trajectories in phase (translational oscillation mode) as $\phi_L = 0$ and $\phi_R = 0$ (dotted grey line in Fig. \ref{comparacionphis}) and $\phi_L = \phi_R = \pi/4$ (solid red line in Fig. \ref{comparacionphis}) in which the behaviour is also exponential as we will see in the following Section. Furthermore, in the blue dashed line we show $\phi_L = 0$ and $\phi_R = \pi/4$ and also $\phi_L = 0$ and $\phi_R = 0.35 \pi$ in the dot-dashed black line. Both examples show an exponential growth in time of the number of particles created with excitation $\Omega = \omega_5$ for mode $n = 3$. 

\subsubsection {Translational modes}

For the particular situation that $\epsilon_R=\epsilon_L = \epsilon$, $\Omega_R=\Omega_L$ and $\phi_L = \phi_R=0$, the cavity oscillates 
as a whole. In this case, the mechanical length is kept constant and is pictorially called ``electromagnetic shaker" \cite{DalvityDiego}. 
As in the previous examples, it is well known that, due to parametric resonance, a naive perturbative solution of Eq. (\ref{ecmov}), in powers of $\epsilon$, breaks down after a short amount of time. In order to find an analytical solution valid for longer times one can use the Multiple Scale Analysis (MSA) technique \cite{crocce,msa}. The MSA provides us with a simple technique equivalent to summing the most secular terms to all order in the perturbative treatment. In this way, it is possible to get a solution valid for a period of time longer than the perturbative case. We shall  introduce a second time scale $\tau = \epsilon t$, and write Eq.(\ref{freemode}) as: 
\begin{equation}Q_n(t,\tau) = \frac{A_n(\tau)}{\sqrt{2 \omega_n}}  e^{-i \omega_n t} + \frac{B_n(\tau)}{\sqrt{2 \omega_n}} 
e^{i \omega_n t}, \label{msamode}
\end{equation}
where the functions $A_n$ and $B_n$ are slowly varying and contain the cumulative resonant effects. To obtain differential equations for them, we insert this ansatz into Eq. (\ref{ecmov}), expand in powers of $\epsilon$ up to first order, and neglect second derivatives of $A_n$ and $B_n$. 
The basic idea of MSA is to impose the condition that any term on the right-hand side of the previous equation with a time dependency of the form $e^{\pm i \omega_nt}$ must vanish. If not, these terms would be in resonance with the left-hand-side term and secularities would appear.  We will follow this procedure in order to get some analytical predictions about particle creation in the translational mode. After imposing the requirement that no term   $e^{\pm i \omega_nt}$ appear, we get:
\begin{eqnarray} \frac{dA_n}{d\tau} &=& \frac{A_0\Omega}{4L_0 \omega_n} \sum_m \omega_m {\tilde b}^{(1)}_{nm} \left\{\left[  \delta(\Omega + \omega_m - \omega_n)  +  \delta(\Omega- \omega_m + \omega_n) \right]A_m -  \left[  \delta(\Omega - \omega_m - \omega_n)  +  \delta(\Omega + \omega_m + \omega_n) \right] B_m\right\} \nonumber \\
&-& \frac{A}{16L_0^2 \omega_n} \sum_m {\tilde g}^{(1)}_{nm}  \left\{\left[  \delta(\Omega + \omega_m - \omega_n)  -  \delta(\Omega- \omega_m + \omega_n) \right]A_m +  \left[  \delta(\Omega - \omega_m - \omega_n)  -  \delta(\Omega + \omega_m + \omega_n) \right] B_m \right\} \label{Adt}
\end{eqnarray}

\begin{eqnarray} \frac{dB_n}{d\tau} &=& - \frac{A_0\Omega}{4L_0 \omega_n} \sum_m \omega_m {\tilde b}^{(1)}_{nm} \left\{\left[  \delta(\Omega + \omega_m + \omega_n)  +  \delta(\Omega- \omega_m - \omega_n) \right]A_m -  \left[  \delta(\Omega - \omega_m + \omega_n)  +  \delta(\Omega + \omega_m - \omega_n) \right] B_m\right\} \nonumber \\
&+& \frac{A}{16L_0^2 \omega_n} \sum_m {\tilde g}^{(1)}_{nm}  \left\{\left[  \delta(\Omega + \omega_m + \omega_n)  -  \delta(\Omega- \omega_m - \omega_n) \right]A_m +  \left[  \delta(\Omega - \omega_m + \omega_n)  -  \delta(\Omega + \omega_m - \omega_n) \right] B_m \right\}, \label{Bdt}
\end{eqnarray}
where $A_0$ is the amplitude of the mirrors displacement and the first order coefficients are given by:

\begin{eqnarray}{\tilde b}^{(1)}_{nm} &=& \frac{4 n m }{n^2 - m^2} \left[(-1)^{n+m} - 1\right] \nonumber \\
{\tilde g}^{(1)}_{nm} &=& - \frac{n m }{n^2 - m^2} 8 L_0 \Omega^2 \left[(-1)^{n+m} - 1\right].
\label{coeflineales}\end{eqnarray}
In these equations, there is not a time-dependent frequency term, from which one can get one resonant mode evolution when exciting the 
system with $\Omega = 2\omega_n$. Therefore, in this case, the perturbation part of the mode 
equation is given in terms of an infinite sum of resonant terms. It is easy to see that, in the small amplitude of the perturbative regime, we will obtain an exponentially in time amount of created particles when $\Omega = \vert \omega_n \pm \omega_m \vert$ (with ``n+m" an odd number from Eq.(\ref{coeflineales})). Therefore, only  
those modes $n$ and $m$ will be parametrically excited \cite{crocce, nosotros}; while other modes will not be excited. In order to set a simple example, we can suppose that the external frequency is given by $\Omega = \omega_2 + \omega_3$. In this case, the linear equations are given by: 
\begin{eqnarray} \ddot A_2 - \left(\frac{6A_0\Omega}{5L_0}\right)^2 \frac{(\omega_3 - \omega_2)^2}{\omega_2 \omega_3} A_2 &=& 0, \\
\ddot B_2 - \left(\frac{6A_0\Omega}{5L_0}\right)^2 \frac{(\omega_3 - \omega_2)^2}{\omega_2 \omega_3} B_2 &=& 0,
\end{eqnarray} 
and the solution for the $B_2$ coefficient is: $B_2(t) \sim \exp( \Gamma \epsilon t)$ where the rate $\Gamma = (6A_0/5L_0)  (\omega_3^2 - \omega_2^2)/\sqrt{\omega_2 \omega_3}$.  This will produce that $N_2$ be exponential with time.  

\begin{figure}[h!]
\begin{minipage}{8.5cm}
\begin{center}
\includegraphics[width=8cm]{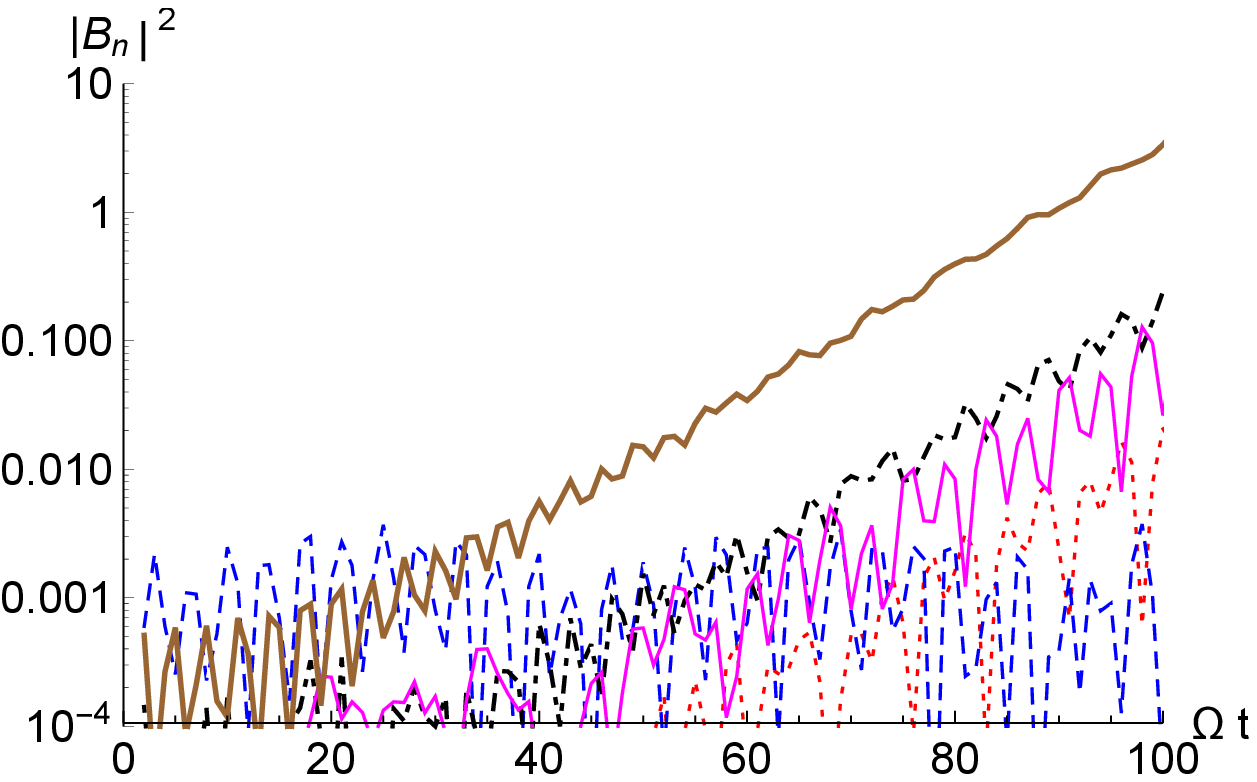}
\end{center}
\end{minipage}
\begin{minipage}{8.5cm}
\begin{center}
\includegraphics[width=8cm]{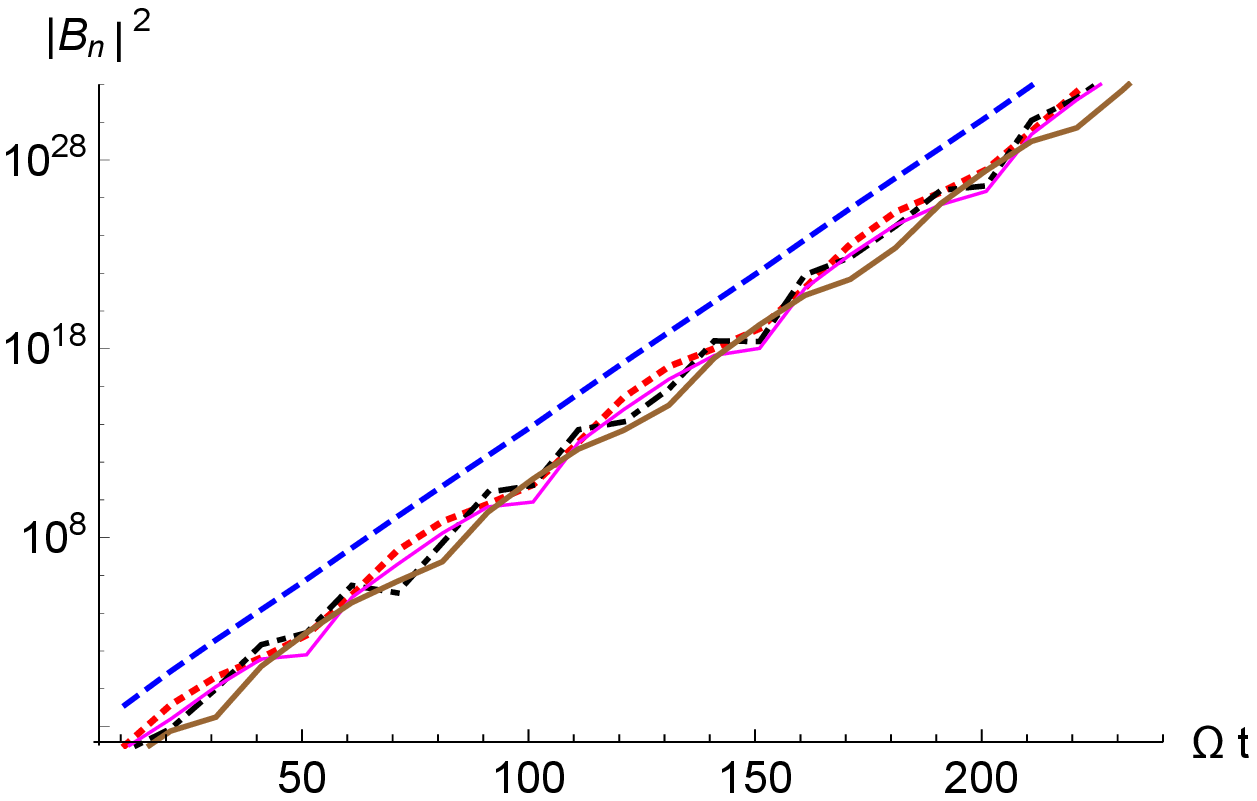}
\end{center}
\end{minipage}
\caption{(Color online) (a) Left: solid lines for $|B_8|^2$ and $|B_5|^2$, the others lines correspond to $|B_3|^2$, $|B_2|^2$ and $|B_1|^2$.  Parameters used: $\epsilon=0.01$, $\Lambda=10$, $M=10$. (b) Right: All modes are parametrically excited for a bigger value of the mass. Parameters used: $\epsilon=0.001$, $\Lambda=10$, $M=50$. Excitation frequency $\Omega = \omega_2 + \omega_3$.}
\label{traslacional2}
\end{figure}
The same conclusion could be obtained for the mode $n= 3$ and that would yield  $N_3$ to be also an exponential function of time.
All other modes different to  $n= 2$ and $n = 3$, are not parametrically excited in the perturbative regime under MSA. 

However, for given values of $\epsilon$ and $M$, the perturbation in the Eqs. (\ref{Adt}) and (\ref{Bdt}) is large, and the perturbative and MSA approaches are not longer valid. Nonetheless, our numerical evidence goes beyond the perturbative regime and the MSA improvement itself and we can study the mode fields at longer times. In  Fig. \ref{traslacional2}, we show an example of non-perturbative result in which we can find an exponentially growing number of created particles for mostly (or all) modes, after exciting the system with $\Omega = \omega_2 + \omega_3$ . This is in principle a non-expected result from the perturbative approach mentioned above.
In Fig.\ref{traslacional2} (a), we can see the behaviour of some field modes for the excitation $\Omega = \omega_2 + \omega_3$, where some modes oscillates while other are already excited for time $\Omega t \sim 1/\epsilon$.  In  Fig.\ref{traslacional2} (b), for a bigger value of mass, we can see that all modes are parametrically excited for the same time scale, even though $\epsilon$ is smaller.  

\subsubsection {Breathing modes}

\begin{figure}[h!]
\begin{minipage}{8.5cm}
\begin{center}
\includegraphics[width=8cm]{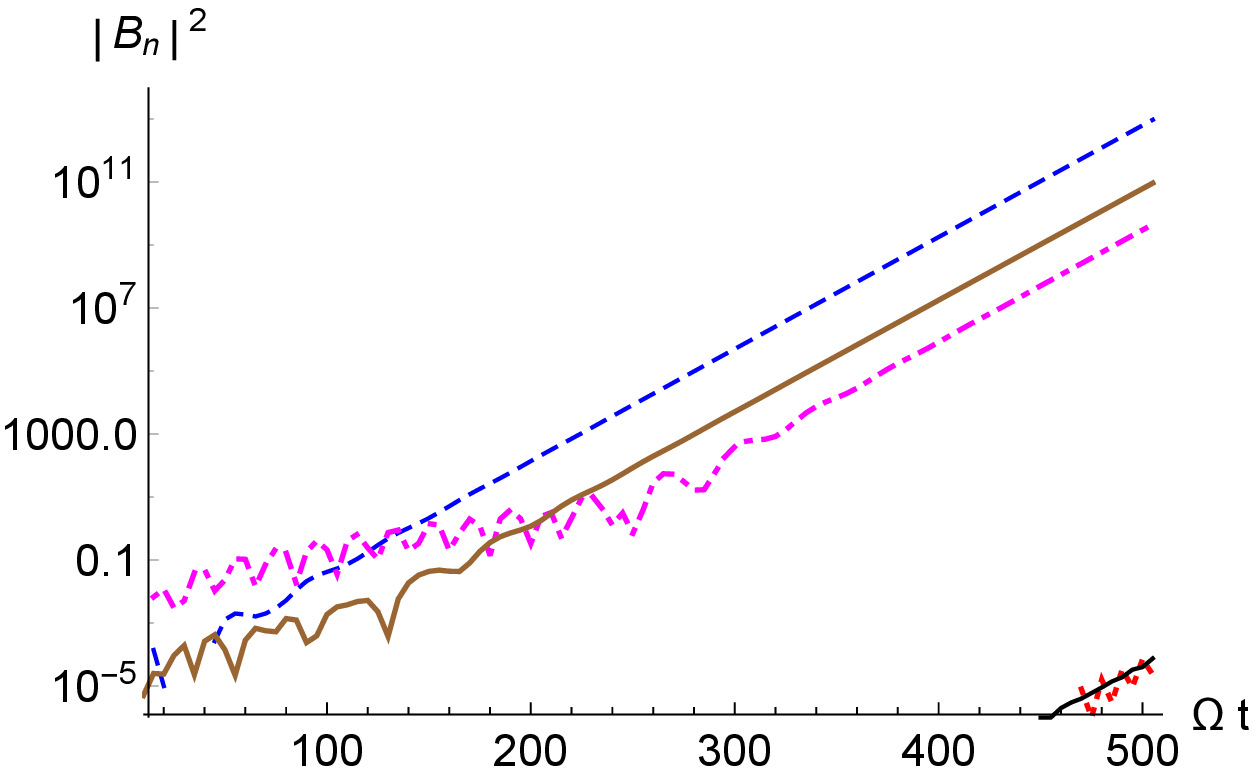}
\end{center}
\end{minipage}
\begin{minipage}{8.5cm}
\begin{center}
\includegraphics[width=8cm]{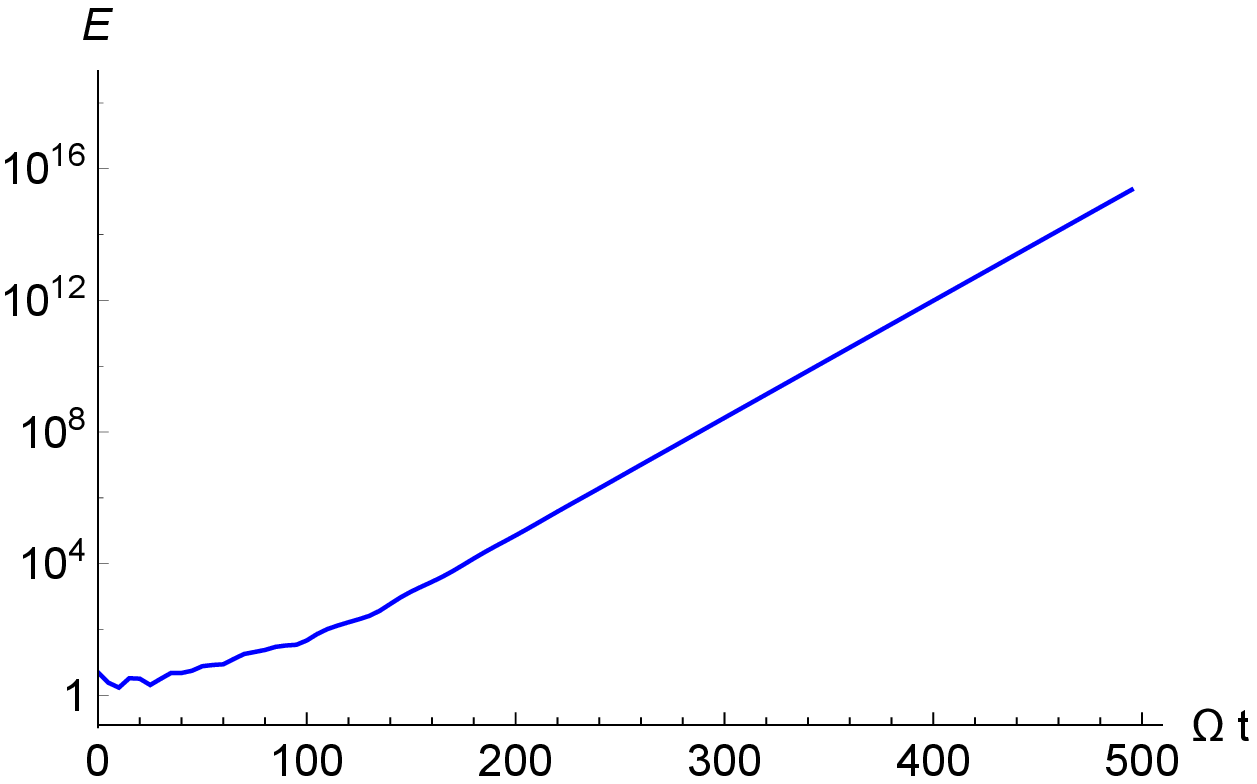}
\end{center}
\end{minipage}
\caption{(Color online) (a) Left: For the breathing modes, we have at left: $\vert B\vert^2$ coefficient versus time for different field modes $n= 2$ ($\vert B_2\vert^2$), $n = 4$ ($\vert B_4\vert^2$), and $ n = 6$ 
($\vert B_6\vert^2$) under $\Omega_R=\Omega_L= \omega_4$ and $\phi_L=\pi$. Field modes $n = 1$ ($\vert B_1\vert^2$) and $n = 3$ ($\vert B_3\vert^2$) seem to get excited at longer times.  (b) Right:  Energy density  as a function of the dimensionless time. Parameters used: $\epsilon=0.01$, $\Lambda=10$. In our units, energy is measeured in units of $1/L_0$.}
\label{caso1phipiq4}
\end{figure}

We next consider the case in which $\epsilon_R=\epsilon_L$, $\Omega_R=\Omega_L= \omega_n$ and $\phi_R=\pi$ ($\phi_L=0$), representing the situation in which the mirrors oscillate symmetrically with respect to the center of the cavity. The mechanical length changes periodically as an ``antishaker", in relation to the previous example in which the cavity moves as a whole. This is simulated in Fig.\ref{caso1phipiq4}. We show the blue dashed line for the mode $n=2$, the dot-dashed magenta line for $n=4$ and the solid line for $n=6$. The odd modes, such as $n=1$ and $n=3$, activate at longer times as shown in Fig.\ref{caso1phipiq4}(a).
In Fig.\ref{caso1phipiq4}, we are exciting the system by $\Omega_R=\Omega_L= \omega_4$ and $\phi_L=\pi$ with a negligible mass, yielding $\omega_4= 4 \pi/L_0$. Therein, we can observe that all even modes are getting excited while the odd modes become excited at longer times (they appeared at times $\Omega t \sim 500$). At the end, all modes become exponentially excited in a non-perturbative example. 

\subsubsection {Three dimensional cavities with two moving mirrors}

\begin{figure}[!h]
\begin{minipage}{8.5cm}
\begin{center}
\includegraphics[width=8cm]{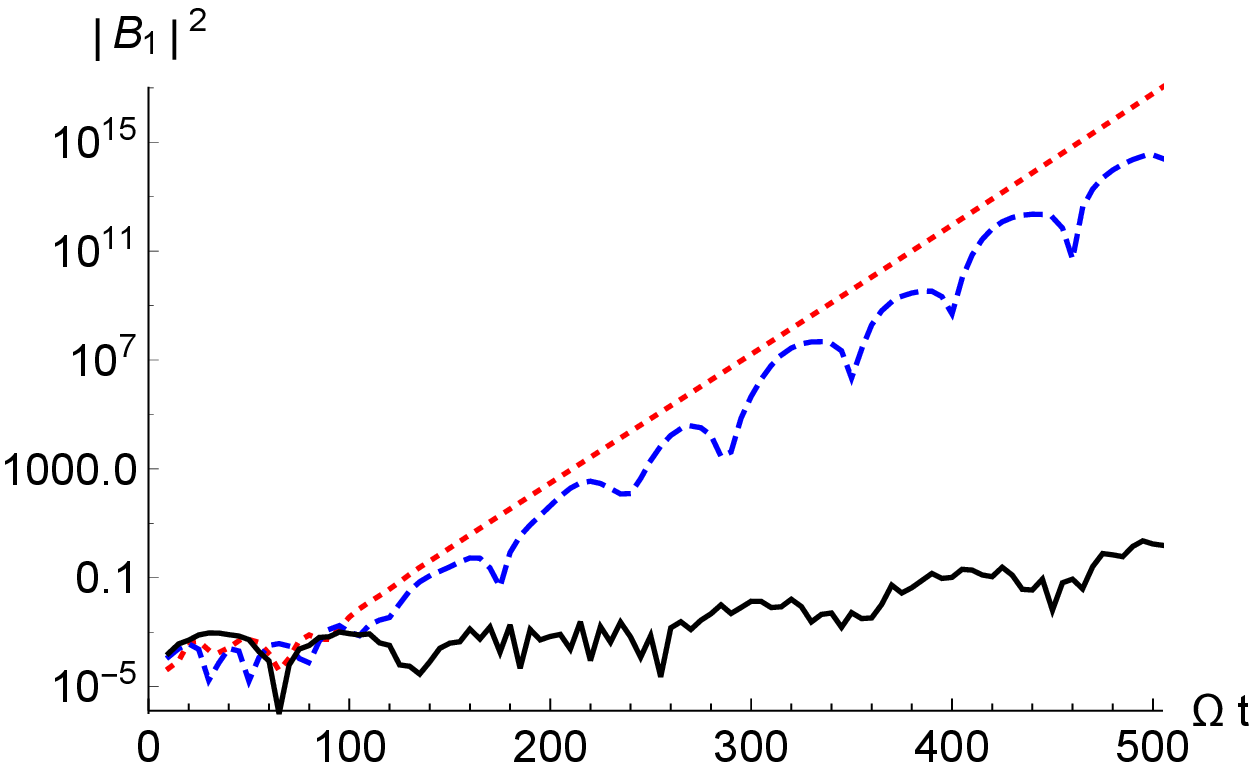}
\end{center}
\end{minipage}
\begin{minipage}{8.5cm}
\begin{center}
\includegraphics[width=8cm]{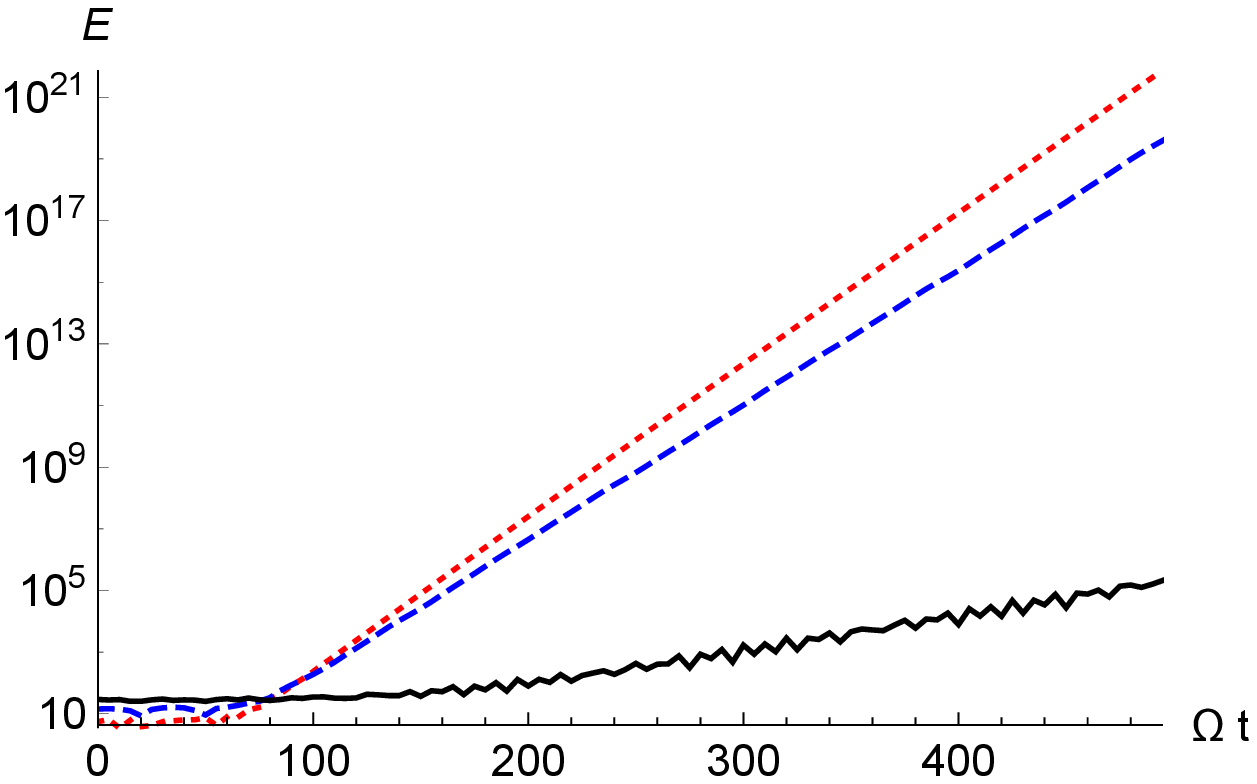}
\end{center}
\end{minipage}
\caption{(Color online) (a) Left: ${B}_1$ coefficient for field mode 1 under $\Omega_R=\Omega_L= \omega_5$ and $\phi_L=0$ for different values of the mass: $M=0.01$ (red dots), $M=1$ (blue dashes) and $M=5$ (black solid curve). (b) Right: Energy as a function of time. Parameters used: $\epsilon=0.01$, $\Lambda=10$. In our units, energy is measeured in units of $1/L_0$.}
\label{tresD}
\end{figure}

We can even study particle creation in the three dimensional cavity with two moving mirrors. Then, we start increasing the value of the mass, and see how the explosive cocktail of Fig. \ref{tresD} starts changing considerably. In Fig.\ref{tresD}, we show the behaviour of the coefficient for field mode $n=1$,  $\vert{B}_1\vert^2$ under  a perturbation defined by $\Omega_R=\Omega_L= \omega_5$ and $\phi_L=0$ for different values of the mass value: $M=0.01$ (dotted line), $M=1$ (dashed) and $M=5$ (solid). In the right side of the figure, we show the energy as a function of time for the same values of the parameters. Therein, we see that for $M=0.01$, the energy has an exponential behaviour while for $M=5$ it  is not.

It is easy to see in Fig.\ref{tresD}  the different results for the number of created particles between one and three dimension cavities.  In fact, it is important to note in Fig. \ref{tresD} (a) that the smaller the mass value, the bigger the growth of the coefficient 
$\vert{B}_1\vert^2$. Then, one dimensional cavities with two oscillating mirrors produce a bigger excitation of modes than the 
corresponding cubic cavity in three dimensions. The same mass hierarchy can be seen in Fig. \ref{tresD} (b) for the energy density inside the cavities (one dimensional cavity when $M\rightarrow 0$, and three dimensional ones for bigger $M$). 

\section{Conclusions}
\label{conclusiones}

In this paper we have presented a detailed numerical analysis of the particle creation for a quantum field in a cavity with two perfectly conducting moving mirrors. This approach was applied to a one dimensional as well as to a three-dimensional cavity box. We have derived the equation of motion of the field modes  and numerically evaluated the Bogolubov transformation between {\it in} and {\it out}-states. From Bogolubov coefficients, we were able to numerically calculate the number of created particles after the mirrors stop moving and return to the 
unperturbed position.

In the case of a one moving wall at $x=R(t)$, while the wall at $x=0$ remains at rest, we have recovered the very well known results. In that case, we have showed that the rate of particle production depends strongly on whether the frequency spectrum is equidistant or not, obtaining the correct  behaviour with respect to the number of created particles. A quadratic behaviour for short timescale and a linear growing for larger temporal scales when the spectrum is equidistant. We also found that, beyond perturbative predictions, the dependance to the number of created particles is exponential with time for very large times. In the case of three-dimensional cavities with one moving mirror, we have shown that our results are in 
agreement with Literature for the perturvatibe or MSA regions, and have provided numerical evidence of the bahaviour of the number of created particles even in the non-perturvatibe case.   

In the case of two moving walls, we have shown that the rate of particle creation depends strongly on the relation among the amplitudes, the frequency and the phase difference in the mirrors' oscillations. We have shown that in some cases there are constructive interference leading to an exponential growth of particles inside the cavity, while for other relations there exist destructive interference with no vacuum radiation. 
We have gone further all analytical studies by considering a massive field and computing the energy density inside the cavity. 

We also study non-perturbative regimes for translational modes obtaining that all modes into the cavity grows exponentially when the amplitude of the perturbation in the mode equation is large (compared with perturbative expansions analysis). Then, we report an exponentially growing number of created particles when exciting with external frequencies $\Omega = \vert \omega_n \pm \omega_m\vert$ (with $n+m$ an odd number). 
We will present a further analysis of the translational mode oscillation with analytical and more numerical support elsewhere \cite{next}. 


\section*{Acknowledgements}
This work was supported by ANPCyT, CONICET, and Universidad de Buenos Aires (UBA). We thanks F.D. Mazzitelli for useful comments.



\begin{thebibliography}{bib}
\bibitem{comentario} For the case of a $3+1$ field, the procedure is similar by assuming  the instantaneous mode basis as\begin{equation}
\phi_k(\vec{x},t)=\sqrt{\frac{2}{R-L}} \sin \bigg(\frac{k_x \pi (x-L)}{R-L} \bigg)\sqrt{\frac{2}{L_y}} \sin \bigg(\frac{k_y \pi y}{L_y} \bigg)\sqrt{\frac{2}{L_z}} \sin \bigg(\frac{k_z \pi z}{L_z} \bigg)\nonumber 
\end{equation}
\end{thebibliography}
\end{document}